# Deep Reinforcement Learning Based Optimal Energy Management of Multi-energy Microgrids with Uncertainties

Yang Cui, *Member, IEEE, Member, CSEE*, Yang Xu, *Student Member, IEEE, Member, CSEE*, Yang Li, *Senior Member, IEEE*, Yijian Wang, Xinpeng Zou

*Abstract*—Multi-energy microgrid (MEMG) offers an effective approach to deal with energy demand diversification and new energy consumption on the consumer side. In MEMG, it is critical to deploy an energy management system (EMS) for efficient utilization of energy and reliable operation of the system. To help EMS formulate optimal dispatching schemes, a deep reinforcement learning (DRL)-based MEMG energy management scheme with renewable energy source (RES) uncertainty is proposed in this paper. To accurately describe the operating state of the MEMG, the off-design performance model of energy conversion devices is considered in scheduling. The nonlinear optimal dispatching model is expressed as a Markov decision process (MDP) and is then addressed by the twin delayed deep deterministic policy gradient (TD3) algorithm. In addition, to accurately describe the uncertainty of RES, the conditional-least squares generative adversarial networks (C-LSGANs) method based on RES forecast power is proposed to construct the scenarios set of RES power generation. The generated data of RES is used for scheduling to obtain caps and floors for the purchase of electricity and natural gas. Based on this, the superior energy supply sector can formulate solutions in advance to tackle the uncertainty of RES. Finally, the simulation analysis demonstrates the validity and superiority of the method.

Nomenclature

*A Parameters*

| | |
|---|---|
| $P_{e,\max}^{CHP}, H_{h,\max}^{GB}$ | Rated capacity of CHP and GB |
| $K_{e,i}^{CHP}, K_{h2e,i}^{CHP}$ | Fitting coefficient of generation efficiency and the ratio of heat to electricity of CHP. |
| $K_i^{GB/EC/AC}$ | Fitting coefficient of the energy conversion efficiency of GB/EC/AC |
| $Q_{c,\max}^{EC/AC}$ | Rated capacity of EC and AC |
| $\lambda^{ESS/TSS/CSS}$ | Self-decay rate of ESS/TSS/CSS. |
| $\eta_{ch/dis}^{ESS/TSS/CSS}$ | Charging/discharging efficiency of ESS/TSS/CSS |
| $C_{\max}^{ESS/TSS/CSS}$ | Rated capacity of ESS/TSS/CSS |
| $L_{cyc}^{80\%}$ | Cycle life of the ESS at SOH=80% |
| $\rho_t^{e/gas}$ | Price of electricity/gas. |
| $\alpha^{CO_2}$ | Carbon tax price |
| $\beta^{e/gas}$ | Carbon intensity of the purchased electricity/gas |
| $P_{e,\max/\min}^{CHP}$ | Upper/lower limits of CHP output power |
| $H_{h,\max/\min}^{GB}$ | Upper/lower limits of GB output power |
| $Q_{c,\max/\min}^{EC}$ | Upper/lower limits of EC output power |
| $Q_{c,\max/\min}^{AC}$ | Upper/lower limits of AC output power |
| $SOC_{e,\max/\min}^{ESS}$ | SOC upper/lower limits of ESS |
| $P_{ch/dis,\max}^{ESS}$ | Charging/ discharging power upper limit of ESS |
| $SOC_{h,\max/\min}^{TSS}$ | SOC upper/lower limit of TSS |
| $H_{ch/dis,\max}^{TSS}$ | Charging/ discharging power upper limit of TSS |
| $SOC_{c,\max/\min}^{CSS}$ | SOC upper/lower limits of CSS |
| $Q_{ch/dis,\max}^{CSS}$ | Charging/ discharging power upper limit of CSS |

*B Variables*

| | |
|---|---|
| $P_{e,t}^{CHP}, H_{h,t}^{CHP}$ | Electricity and thermal energy generated of CHP |
| $G_{g,t}^{CHP}$ | Natural gas consumed of CHP |
| $\eta_{e,t}^{CHP}, \eta_{h2e,t}^{CHP}$ | Electricity efficiency and the ratio of heat to electricity of CHP |
| $H_{e,t}^{GB}, G_{g,t}^{GB}$ | Thermal generated and natural gas consumed of GB |
| $\eta_t^{GB/EC/AC}$ | Energy conversion efficiency of GB/EC/AC |
| $Q_{c,t}^{EC}, P_{e,t}^{EC}$ | Cool generated and electricity consumed of EC |
| $Q_{c,t}^{AC}, H_{h,t}^{AC}$ | Cool generated and thermal energy consumed of AC |
| $SOC_t^{ESS/TSS/CSS}$ | SOC of ESS/TSS/CSS |
| $P_{ch/dis,t}^{ESS}$ | Charging/ discharging power of ESS |
| $H_{ch/dis,t}^{TSS}$ | Charging/ discharging power of TSS |
| $Q_{ch/dis,t}^{CSS}$ | Charging/ discharging power of CSS |
| $P_t^{grid}, G_t^{grid}$ | Purchased electrical energy and natural gas |
| $P_{e,t}^{WT}, P_{e,t}^{PV}$ | Wind and photovoltaic power output |

## I. INTRODUCTION

To cope with the crisis of energy and the emissions of greenhouse gases, renewable energy sources (RES) are receiving widespread attention [1]. However, the natural characteristics of RES such as randomness and volatility will

Y. Cui, Y. Xu (corresponding author, e-mail: xy961223@126.com), Y. Li, Y. J. Wang and X. P. Zou are with Northeast Electric Power University, Jilin, 132012, China.

affect the reliable operation of the grid, which also inhibits the development of RES [2]. In addition, individual small-capacity distributed energy resources (DERs) cannot participate directly in the system operation of the superior grid because of the minimum admission capacity issued by the power sector manager [3]. Therefore, the microgrid technology not only helps to solve the above problems but also can realize the local consumption of RES, which reduces the loss of transmission process. With the diversification of consumer energy consumption, single-energy microgrids (SEMG) [4-5] are transforming into MEMG [6-7]. Therefore, with the continuous promotion of MEMG, how to achieve the optimal energy management of MEMG is an essential research topic.

At the MG level, a MEMG is composed of RES, energy conversion equipment, energy storage systems, and local energy consumption [8]. As for solving optimal energy management problems, there are four main categories of methods: mathematical programming (MP), control theory-based, metaheuristic optimization-based, and machine learning-based.

To obtain optimal scheduling strategy, the MP method has been widely employed in MEMG for many years [9-13]. In [14], the stochastic optimal scheduling model of isolated MG was proposed based on RES and load forecast data. A MEMG, incorporating energy conversion equipment and energy storage systems, was detailed in [15]. After the establishment of the mathematical model, solving the MP method can be done directly by using a commercial solver. However, when faced with optimization problems featuring nonlinear objective functions or constraints, MP may struggle. Metaheuristic methods are effective measures to deal with nonlinear optimization problems [16-19]. Ref. [20] used the evolutionary particle swarm optimization (PSO) algorithm to tackle dispatch tasks in the system with RES. Ref. [21] uses PSO for optimization problems containing non-linear constraints. Ref. [22] proposed a memory-based genetic algorithm (GA) to formulate generation tasks for MG and the superiority of the memory-based GA was verified in simulation. However, the metaheuristic methods easily result in poor search efficiency due to the past experience cannot be reused for new tasks. In addition, control-based methods are also applied to optimize energy management [23-24]. Model predictive control (MPC) was employed to optimally manage the working of the cluster MG in [25]. In [26], a stochastic MPC model is used for the scheduling and control of MGs for large-scale RES. Ref. [27] used approximate dynamic programming to deal with the Markov decision process (MDP).

The above method is considered to be a model-based method. As the optimization variables in MG increase, the calculation cost will also increase. Furthermore, disruptions to the system necessitate a complete re-optimization for both MP and metaheuristic methods, exacerbating the computational burden. Therefore, with the boom in machine learning, model-free deep reinforcement learning (DRL) algorithms have been proposed [28]. DRL is a data-driven approach that has been widely used to optimal decision-making problems in energy fields, such as electricity market equilibrium [29], electric vehicle charging strategy [30-32], and wind speed forecasting [33-34]. In addition, there has been considerable success in optimizing energy management. For instance, distributed proximal policy optimization (PPO) was employed to obtain the dispatching plans of the SEMG in [35]. Ref. [36] proposed a Q-learning based approach to optimize the energy distribution of the MG. An improved soft actor critic (SAC) was applied in [37] to manage the energy distribution of multi-energy systems, and the simulation results indicate that the proposed method can improve the economy of the system. Ref. [38] applied rainbow Deep Q-Networks to manage a battery for energy arbitrage. An improved deep deterministic policy gradient (DDPG) algorithm was applied in [39] to obtain a dispatch scheme. Ref. [40] applied the PPO algorithm to solve dynamic scheduling tasks for IES, which can reduce the system operator's operating cost. Yet, a notable limitation in these methods [37, 39-40] is the oversimplification of the energy conversion relationship, often linearized. In actuality, the energy conversion device usually has obvious off-design performance, that is, the output characteristics of the device under different environments and load rates are obviously different. The linear operation model ignores or simplifies the off-design performance of devices, which will cause the deviation of the energy conversion relationship of the system to a certain extent. To more accurately describe the input and output relationship of energy conversion devices, the off-design performance of devices is necessary to be considered.

The uncertainty of RES forecast power will affect the economy of scheduling and the stability of the system. Approaches to solving uncertainty problems fall into four major categories: robust optimization (RO) [12] and stochastic programming (SP) [14], interval programming (IP) [41], and fuzzy chance-constrained programming (FCCP) [42]. IP and FCCP are suitable for scenarios with insufficient historical data. However, their uncertainty modeling is crude in MG with large amounts of data. RO adopts uncertainty sets to model uncertainties and it obtains a scheduling strategy under the worst-case situation, which results in the conservativeness of scheduling results and poor economic advantages. SP utilizes the scenario-based approach to model uncertainty, which focuses on expected costs and can decrease conservativeness. To describe the uncertainty of RES, the traditional approach is usually modeled by statistical methods, where the specific parameters of the model are solved by historical data [43]. However, it is not easy to obtain an exact probability distribution. Hence, Ref. [44] proposed a model-free renewable scenario generation method using generative adversarial networks (GANs). For the privacy protection of multiple subjects, a renewable scenario generation method combining federated learning and least square GANs (LSGANs) is proposed in [45]. Although the RES data with the full diversity of behaviors can be generated in [44-45], they cannot all be applied to scheduling. Hence, it is necessary to utilize forecast information to generate data under specific conditions.

To address the shortcomings of the above studies, this paper proposes an energy management strategy of MEMG considering both the off-design performance of devices and the uncertainty of RES. The key elements of this study are described in summary as follows:

1) To accurately describe the operating state of the system, the off-design performance model of energy conversion devices is considered. The nonlinear optimal model of MEMG with the off-design performance of devices is expressed as

MDP, which is processed using DRL. This is also the first work to use TD3 to deal with the nonlinear optimization problem.

2) A novel renewable scenario generation approach based on the forecast power of RES is proposed. To the best of the authors' knowledge, this is the first work that C-LSGANs used to generate scenario sets to describe the uncertainty of RES.

3) The generated data of RES is used for scheduling to obtain the caps and floors of purchased energy. Based on this, the superior energy supply sector can formulate solutions in advance to guarantee the reliable operation of the MEMG.

The remaining parts of this study are described below. Section II presents the operation model of MEMG devices. Section III establishes the optimized scheduling model of MEMG considering the off-design performance of devices and formulates it as MDP. C-LSGANs based RES scenario generation method is shown in Section IV. Section V performs the simulation analysis and concludes with the conclusion.

## II. OPERATION MODEL OF MEMG DEVICES

With the rapid development of RES and the diversification of consumer energy consumption, the MEMG system is a key technology to solve this problem. A typical MEMG structure is shown in Fig. 1. The energy supply unit consists of the superior power grid, the superior gas grid, and RES. The energy conversion units are composed of combined heat and power (CHP), electric chiller (EC), gas boiler (GB), and absorption chillers (AC). The energy storage unit includes electrical storage systems (ESS), cold storage systems (CSS), and thermal storage systems (TSS). The energy consumption unit consists of power energy consumption, heat energy consumption, and cold energy consumption.

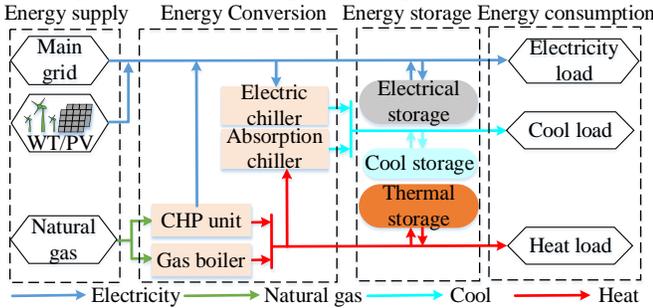

Fig. 1. The structure of MEMG

### A. Model of energy conversion devices
*1) Off-design performance of CHP*

The CHP unit, as a high-efficiency energy conversion device, has been widely installed. It consumes natural gas to generate electrical energy and thermal energy, allowing for the local consumption of energy. Hence, energy losses during transmission can be avoided. In addition, CHP can also reduce emissions of greenhouse gases and other air pollutants due to the use of natural gas as a fuel. The many advantages of CHP such as high efficiency, environmental protection, and economics, make it very popular. The energy conversion model of CHP is shown as follows:

$$\begin{cases} P_{e,t}^{CHP} = \eta_{e,t}^{CHP} G_{g,t}^{CHP} \\ H_{h,t}^{CHP} = \eta_{h2e,t}^{CHP} P_{e,t}^{CHP} \end{cases} \quad (1)$$

However, CHP will not always operate under rated conditions during actual operation. Both its electricity efficiency and the ratio of heat to electricity have a certain relationship with the load rate. The off-design performance model of CHP based on the part-load characteristics can be expressed in (2).

$$\begin{cases} \eta_{e,t}^{CHP} = \sum_{i=0}^{n} K_{e,i}^{CHP} \left( \frac{P_{e,t}^{CHP}}{P_{e,max}^{CHP}} \right)^i \\ \eta_{h2e,t}^{CHP} = \sum_{i=0}^{n} K_{h2e,i}^{CHP} \left( \frac{P_{e,t}^{CHP}}{P_{e,max}^{CHP}} \right)^i \end{cases} \quad (2)$$

*2) Off-design performance of GB*

GB produces thermal energy by burning natural gas, and because of its convenient installation is widely used in heating places. The combustion efficiency of GB is affected by boiler water temperature, burner working time, and boiler load rate. At different load rates, the energy conversion efficiency of GB is different. The relationship between energy conversion efficiency and the load rate of GB can be expressed as

$$\begin{cases} H_{h,t}^{GB} = \eta_t^{GB} G_{g,t}^{GB} \\ \eta_t^{GB} = \sum_{i=0}^{n} K_i^{GB} \left( \frac{H_{h,t}^{GB}}{H_{h,max}^{GB}} \right)^i \end{cases} \quad (3)$$

*3) Off-design performance of EC*

Commercial and industrial buildings as well as new residential users have widely adopted EC to generate cold energy nowadays. The EC is composed of a condenser, expansion valve, evaporator, and screw compressor. The working principle of EC is shown in Fig. 2. The refrigerant of EC absorbs heat in the evaporator and dissipates heat in the condenser, which completes the cycle of refrigeration.

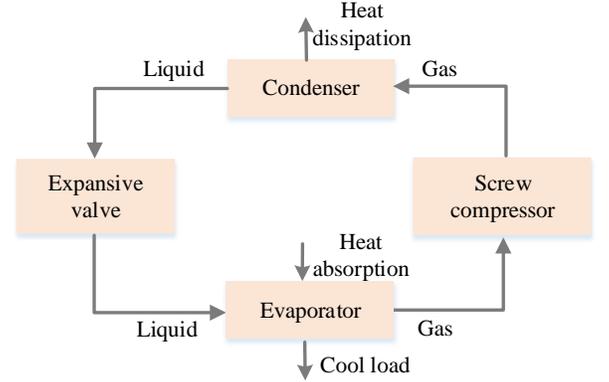

Fig. 2 The working principle of EC

When considering the part-load characteristics of EC, the off-design performance model of EC can be expressed by (4).

$$\begin{cases} Q_{c,t}^{EC} = \eta_t^{EC} P_{e,t}^{EC} \\ \eta_t^{EC} = \sum_{i=0}^{n} K_i^{EC} \left( \frac{Q_{c,t}^{EC}}{Q_{c,max}^{EC}} \right)^i \end{cases} \quad (4)$$

*4) Off-design performance of AC*

AC is a kind of refrigeration equipment using low-temperature heat sources such as flue gas waste heat, waste heat, and low-temperature hot water. The advantage of AC is that it can use low-quality heat energy to generate cold energy with less power energy consumption. The AC is composed of a generator, condenser, evaporator, absorber, solution heat exchanger, and throttle valve. Its working principle is shown in Fig. 3.

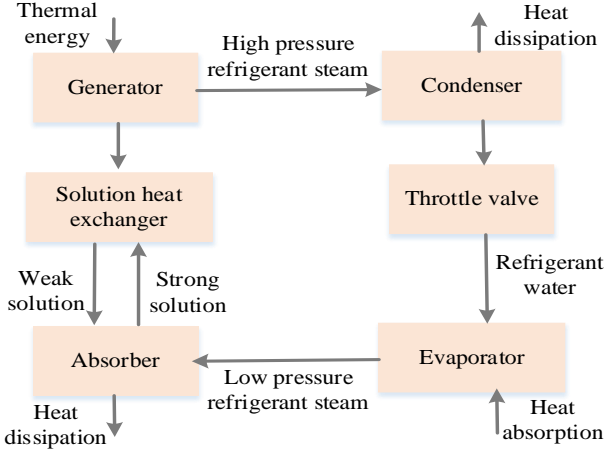

Fig. 3 The working principle of AC

When considering the part-load characteristics of AC, the off-design performance model of the AC can be expressed as follows:

$$\begin{cases} Q_{c,t}^{AC} = \eta_t^{AC} H_{h,t}^{AC} \\ \eta_t^{AC} = \sum_{i=0}^{n} K_i^{AC} \left( \frac{Q_{c,t}^{AC}}{Q_{c,max}^{AC}} \right)^i \end{cases} \quad (5)$$

### B. Model of energy storage devices
#### 1) Model of ESS

The state of charging (SOC) of ESS is a significant parameter, which affects the charging power of ESS at the next moment. The SOC of ESS at time $t$ can be shown in (6).

$$SOC_t^{ESS} = \lambda^{ESS} SOC_{t-1}^{ESS} + \left[ \eta_{ch}^{ESS} P_{ch,t}^{ESS} - \frac{P_{dis,t}^{ESS}}{\eta_{dis}^{ESS}} \right] \Delta t / C_{max}^{ESS} \quad (6)$$

In addition, due to the operation of ESS, battery degradation is inevitable. Therefore, the state of health (SOH) of ESS is introduced for describing the changes in SOC of ESS. A detailed description of SOH can be found in [46].

$$SOH_t^{ESS} = SOH_{t-1}^{ESS} + \left( SOH_{initial}^{ESS} - SOH_{end}^{ESS} \right) \cdot AGE_{total}^{ESS} \quad (7)$$

The degree of overall aging, $AGE_{total,t}^{ESS}$, includes two parts: calendric aging, $AGE_{cal,t}^{ESS}$, and cyclic aging, $AGE_{cyc,t}^{ESS}$. They can be represented in (8) [47].

$$\begin{cases} AGE_{total,t}^{ESS} = AGE_{cal,t}^{ESS} + AGE_{cyc,t}^{ESS} \\ AGE_{cal,t}^{ESS} = 6.6148 \times 10^{-6} \cdot SOC_t^{ESS} + 4.6404 \times 10^{-6} \\ AGE_{cyc,t}^{ESS} = 0.5 \times \frac{\left| \eta_{ch}^{ESS} P_{ch,t}^{ESS} - P_{dis,t}^{ESS} / \eta_{dis}^{ESS} \right| \times \Delta t}{L_{cyc}^{80\%} \times C_{max}^{ESS}} \end{cases} \quad (8)$$

#### 2) Model of TSS
The SOC of TSS at time $t$ can be expressed in (9).

$$SOC_t^{TSS} = \lambda^{TSS} SOC_{t-1}^{TSS} + \left[ \eta_{ch}^{TSS} H_{ch,t}^{TSS} - \frac{H_{dis,t}^{TSS}}{\eta_{dis}^{TSS}} \right] \Delta t / C_{max}^{TSS} \quad (9)$$

#### 3) Model of CSS
The model of CSS is similar to the model of TSS, which can be expressed in (10):

$$SOC_t^{CSS} = \lambda^{CSS} SOC_{t-1}^{CSS} + \left[ \eta_{ch}^{CSS} Q_{ch,t}^{CSS} - \frac{Q_{dis,t}^{CSS}}{\eta_{dis}^{CSS}} \right] \Delta t / C_{max}^{CSS} \quad (10)$$

## III. THE PROBLEM FORMULATION

### A. Mathematical model of optimal energy management
#### 1) Objective function

The MEMG optimal energy management problem is to minimize the system operating cost, which includes the energy purchase cost and the environmental cost. The total operating cost of the system, $C^{Total}$, is represented as

$$C^{Total} = \min(C^{Energy} + C^{Carbon}) \quad (11)$$

The cost of purchased energy, $C^{Energy}$, is as follows:

$$C^{Energy} = \sum_{t=1}^{T} \left( \rho_t^e P_t^{grid} + \rho_t^{gas} G_t^{grid} \right) \quad (12)$$

The environmental cost, $C^{Carbon}$, is the penalty for $CO_2$ emissions caused by purchasing energy from the main grid.

$$C^{Carbon} = \alpha^{CO_2} \sum_{t=1}^{T} \left( \beta^e P_t^{grid} + \beta^{gas} G_t^{grid} \right) \quad (13)$$

#### 2) Constraints
(a) Energy balance constraint:

Electric energy, cold energy, thermal energy, and natural gas need to maintain the balance of supply and demand.

$$P_{e,t}^{grid} + P_{e,t}^{WT} + P_{e,t}^{PV} + P_{e,t}^{CHP} + P_{dis,t}^{ESS} = P_{e,t}^{L} + P_{ch,t}^{ESS} + P_{e,t}^{EC} \quad (14)$$

$$H_{h,t}^{CHP} + H_{h,t}^{GB} + H_{dis,t}^{TSS} = H_{h,t}^{L} + H_{ch,t}^{TSS} + H_{h,t}^{AC} \quad (15)$$

$$Q_{c,t}^{EC} + Q_{c,t}^{AC} + Q_{dis,t}^{CSS} = Q_{c,t}^{L} + Q_{ch,t}^{CSS} \quad (16)$$

$$G_{g,t}^{grid} = G_{g,t}^{CHP} + G_{g,t}^{GB} \quad (17)$$

where $P_{e,t}^{L}$, $H_{h,t}^{L}$, and $Q_{c,t}^{L}$ denotes the demand of electric, thermal, and cold energy, respectively.

(b) Devices operating constraints

The energy conversion device needs to meet the operating limit of the device with the following constraints.

$$P_{e,min}^{CHP} \leq P_{e,t}^{CHP} \leq P_{e,max}^{CHP} \quad (18)$$

$$H_{h,min}^{GB} \leq H_{h,t}^{GB} \leq H_{h,max}^{GB} \quad (19)$$

$$Q_{c,min}^{EC} \leq Q_{c,t}^{EC} \leq Q_{c,max}^{EC} \quad (20)$$

$$Q_{c,min}^{AC} \leq Q_{c,t}^{AC} \leq Q_{c,max}^{AC} \quad (21)$$

The energy storage device must meet the limit of the capacity of the device.

$$\begin{cases} SOC_{e,min}^{ESS} \cdot SOH_t^{ESS} \leq SOC_{e,t}^{ESS} \leq SOC_{e,max}^{ESS} \cdot SOH_t^{ESS} \\ 0 \leq P_{ch,t}^{ESS} \leq P_{ch,max}^{ESS} \\ 0 \leq P_{dis,t}^{ESS} \leq P_{dis,max}^{ESS} \end{cases} \quad (22)$$

$$\begin{cases} SOC_{h,min}^{TSS} \leq SOC_{h,t}^{TSS} \leq SOC_{h,max}^{TSS} \\ 0 \leq H_{ch,t}^{TSS} \leq H_{ch,max}^{TSS} \\ 0 \leq H_{dis,t}^{TSS} \leq H_{dis,max}^{TSS} \end{cases} \quad (23)$$

$$\begin{cases} SOC_{c,min}^{CSS} \leq SOC_{c,t}^{CSS} \leq SOC_{c,max}^{CSS} \\ 0 \leq Q_{ch,t}^{CSS} \leq Q_{ch,max}^{CSS} \\ 0 \leq Q_{dis,t}^{CSS} \leq Q_{dis,max}^{CSS} \end{cases} \quad (24)$$

## B. Reinforcement learning framework of optimal energy management

Reinforcement learning performs adaptive learning through agent "trial and error". The agent interacts with the outside continuously. The maximum cumulative reward is obtained in continuous learning by acquiring the external state, taking actions to change the external state, and receiving a corresponding reward or penalty as a guide for updating the model parameters. The environmental interaction can be generally described by MDP in Fig. 4.

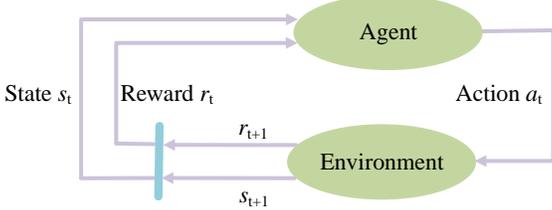

Fig. 4. Agent learning process

The optimal energy distribution of MEMG considering the off-design performance of devices is a nonlinear optimization problem. The problem being addressed can be described as MDP for minimizing the operation cost of MEMG.

a) System states: The system state is defined as the information of the agent obtained from the MEMG, which can be used to reflect the state of the MEMG. The states can be described as

$$s_t = \left( P_t^{WT}, P_t^{PV}, P_{e,t}^{L}, H_{h,t}^{L}, Q_{c,t}^{L}, \rho_t^e, SOC_{e,t-1}^{ESS}, SOC_{h,t-1}^{TSS}, SOC_{c,t-1}^{CSS} \right) \quad (25)$$

b) System actions: The system actions represent scheduling decisions taken by MEMG. The actions can be described as

$$a_t = \left( P_{e,t}^{CHP}, H_{h,t}^{GB}, Q_{c,t}^{EC}, Q_{c,t}^{AC}, P_{e,t}^{grid} \right) \quad (26)$$

c) Reward functions: Based on the states and actions of the MEMG, the system reward function can be obtained. The reward functions are described as follows:

$$r_t = -\left( C_t^{Total} + C_t^{Penalty} \right) \quad (27)$$

In addition, the penalty cost will occur when the energy supply and demand balance is not satisfied.

$$C_t^{Penalty} = \alpha_e \left| P_{e,t}^{S} - P_{e,t}^{L} \right| + \alpha_h \left| H_{h,t}^{S} - H_{h,t}^{L} \right| + \alpha_c \left| Q_{c,t}^{S} - Q_{c,t}^{L} \right| \quad (28)$$

where $\alpha_e$, $\alpha_h$ and $\alpha_c$ are the penalty factor. $P_{e,t}^{S}$, $H_{h,t}^{S}$ and $Q_{c,t}^{S}$ represent the energy supply. $P_{e,t}^{L}$, $H_{h,t}^{L}$ and $Q_{c,t}^{L}$ represent the energy demand.

## C. Solving method of optimal energy management

The goal of the agent in DRL is to search for the optimal control strategy that maximizes the expectation of the cumulative discount reward.

$$\pi^* = \arg\max_{\pi} \mathbb{E}_{a_t \sim \pi(\cdot|s_t)} \sum_{t=0}^{\infty} \gamma^t \left[ r_t(s_t, a_t) \right] \quad (29)$$

where $\pi(a_t | s_t)$ is the policy for the agent. $\gamma$ is the discounting coefficient.

To solve the overestimation and high variance problems lied in deep Q-learning and DDPG algorithms, TD3 is proposed in [48]. The TD3-based scheduling framework constructed is shown in Fig. 5.

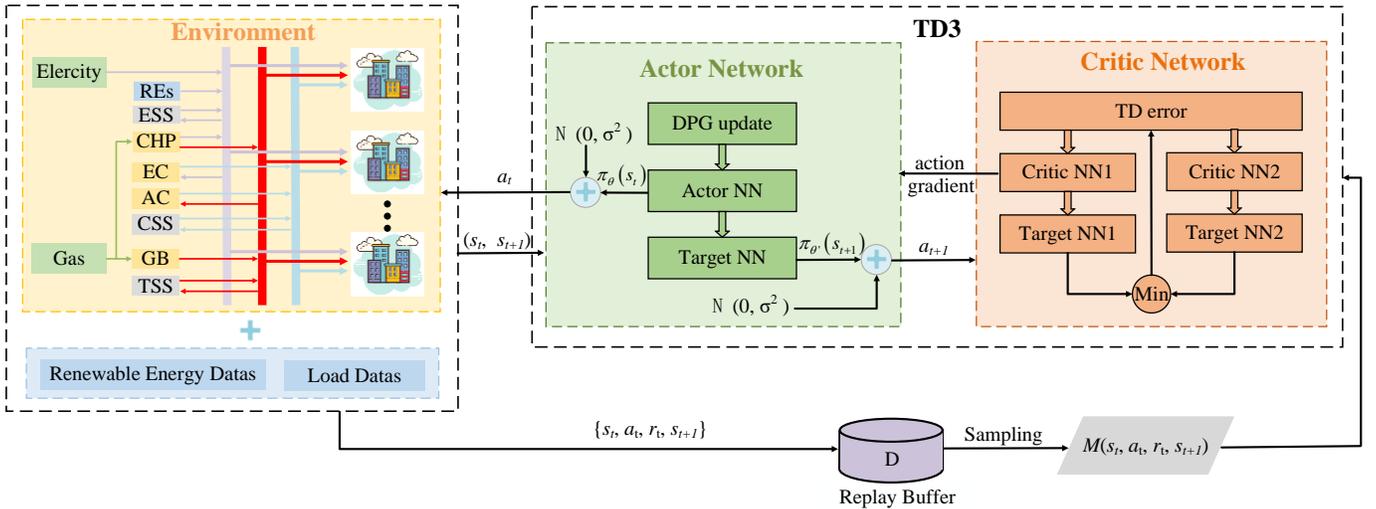

Fig. 5. Interaction framework between agent and environment

In addition, the TD3 update process is as follows:

a) The Q network update

For the Q network in critic, the parameters are optimized by minimizing the loss function $L(\omega)$.

$$L(\omega) = \mathbb{E}_{s_t \sim p_\pi, a_t \sim \pi, r_t \sim R} \left[ \left( Q_\omega(s_t, a_t) - y_t \right)^2 \right] \quad (30)$$

where $p_\pi$ is the state distribution, $\pi$ and $R$ are the distribution of the policy and reward function, respectively, and $y_t$ is the target value.

$$y_t = r_t + \gamma \min_{j \in \{1,2\}} Q_{\hat{\omega}_j}(s_{t+1}, \hat{a}_{t+1}) \quad (31)$$

$$\hat{a}_{t+1} = \pi_{\hat{\theta}}(s_{t+1}) + \tilde{\varepsilon}, \tilde{\varepsilon} \sim \text{clip}(N(0,\sigma), -c, c) \quad (32)$$

where $\hat{a}_{t+1}$ denotes the target action after the clip.

The gradient of $L(\omega)$ to $\omega$ is shown as follows:

$$\nabla_\omega L(\omega) = \mathbf{E}_{s_t \sim p_\pi, a_t \sim \pi, r_t \sim R}\left[2(Q_\omega(s_t, a_t) - y_t)\nabla_\omega Q_\omega(s_t, a_t)\right] \quad (33)$$

On the basis of the gradient rule, the values of the critic network are being adjusted as

$$\omega \leftarrow \omega - \alpha_Q \nabla_\omega L(\omega) \quad (34)$$

b) The policy network update

For the deterministic policy network in the actor, the values are optimized by the sampled policy gradient, which can be presented in (35).

$$\nabla_\theta J \approx \mathbf{E}_{s_t \sim p_\pi}\left[\nabla_a Q_\omega(s, a)|_{s=s_t, a=\pi_\theta(s_t)} \nabla_\theta \pi_\theta(s)|_{s=s_t}\right] \quad (35)$$

On the basis of the deterministic policy gradient, the values of the actor network are being adjusted as

$$\theta \leftarrow \theta + \alpha_\pi \nabla_\theta J \quad (36)$$

c) The target network update

The values of the target network are gently updated, which enhances the stabilization of the learning process.

$$\hat{\omega}_1 \leftarrow \tau\omega_1 + (1-\tau)\hat{\omega}_1 \quad (37)$$

$$\hat{\omega}_2 \leftarrow \tau\omega_2 + (1-\tau)\hat{\omega}_2 \quad (38)$$

$$\hat{\theta} \leftarrow \tau\theta + (1-\tau)\hat{\theta} \quad (39)$$

where $\tau$ is the soft update parameter.

## IV. C-LSGANs BASED RENEWABLE ENERGY SCENARIO GENERATION

### A. Scenario generation method based on C-LSGANs

1) GANs Model

GANs is an unsupervised model proposed by Goodfellow in 2014 [49], which is composed of 2 independent neural networks: generator (G) and discriminator (D). The basic structure of GANs can be shown in Fig. 6.

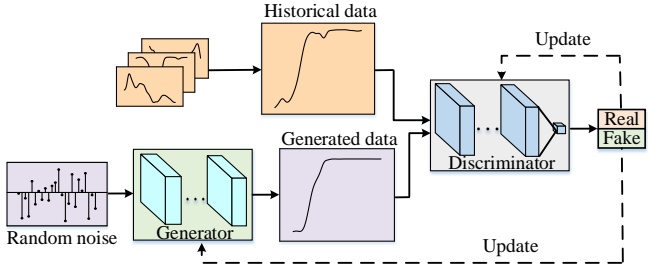

Fig. 6. The learning process of GANs

In GANs, $x$ denotes the actual data samples. $p_{\text{data}}(x)$ represents the actual data distribution. z represents the random noise data. $p_z(z)$ represents the distribution of noise data. For the duration of training, a batch of $z \sim p_z(z)$ is delivered to G to get the generated sample $G(z)$, which is subject to probability distribution $p_g(z)$. The training objective of the G is to make $p_g(z)$ as identical as possible to $p_{\text{data}}(x)$.

The loss functions, $L_G$, and, $L_D$, of G and D can be defined as follows:

$$L_G = \mathbf{E}_{z \sim p_z}\left[\log(1 - D(G(z)))\right] \quad (40)$$

$$L_D = -\mathbf{E}_{x \sim p_{\text{data}}}\left[\log D(x)\right] - \mathbf{E}_{z \sim p_z}\left[\log(1 - D(G(z)))\right] \quad (41)$$

where $\mathbf{E}$ represents the expected value of the data.

Based on (40) and (41), the objective function of GANs is designed as shown in (42):

$$\min_G \max_D V_{\text{GANs}}(D, G) = \mathbf{E}_{x \sim p_{\text{data}}}\left[\log(D(x))\right] + \mathbf{E}_{z \sim p_z}\left[\log(1 - D(G(z)))\right] \quad (42)$$

2) C-LSGANs Model

However, the loss function of the D in the traditional GANs was comparable to the Jensen-Shannon divergence (JSD) [45]. JSD is always very close to a constant when the generated sample distribution and the true sample distribution cannot overlap in all dimensions. At this time, the JSD reaches saturation and the loss of the D is zero. Ultimately, as a result of employing the cross-entropy loss function, conventional GANs suffer from the problem of gradient disappearance in backpropagation, making network training difficult and mode collapse.

To solve the above problem, LSGANs was proposed [50]. The goal function of LSGANs can be shown in (43):

$$\begin{cases} \min_D V_{\text{LSGANs}}(D) = \frac{1}{2}\mathbf{E}_{x \sim p_{\text{data}}}\left[(D(x) - b)^2\right] \\ \qquad + \frac{1}{2}\mathbf{E}_{z \sim p_z}\left[(D(G(z)) - a)^2\right] \\ \min_G V_{\text{LSGANs}}(G) = \frac{1}{2}\mathbf{E}_{z \sim p_z}\left[(D(G(z)) - c)^2\right] \end{cases} \quad (43)$$

where $a$ and $b$ represent labels of generated data and real data, respectively. c represents the value which the G wants the D to trust for generated data. In this work, we set $a=0$, $b=c=1$.

The least squares loss function is used by the D and G in LSGANs to provide more gradients by penalizing samples away from the decision boundary, which alleviates the problem of gradual gradient disappearance. At the same time, the loss function facilitates the generation of higher quality samples since it will penalize the fake samples and "pull" them to the decision boundary.

Furthermore, to generate data under specific events, the conditional information c is added in the generator and discriminator of LSGANs. Hence, the objective function of C-LSGANs is constructed as

$$\begin{cases} \min_D V_{\text{LSGANs}}(D) = \frac{1}{2}\mathbf{E}_{x \sim p_{\text{data}}}\left[(D(x|c) - 1)^2\right] \\ \qquad + \frac{1}{2}\mathbf{E}_{z \sim p_z}\left[(D(G(z|c)))^2\right] \\ \min_G V_{\text{LSGANs}}(G) = \frac{1}{2}\mathbf{E}_{z \sim p_z}\left[(D(G(z|c)) - 1)^2\right] \end{cases} \quad (44)$$

3) C-LSGANs Configuration

In the study, the predicted data are used as conditional information, and the predicted and actual data are sent to C-LSGANs to train. After training, the G of C-LSGANs can learn the mapping relationship between noise distribution and output distribution that satisfies the conditional information. To make full advantage of the forecast information of RES, the deep neural network architecture of the G and D of C-LSGANs was designed in Fig. 7.

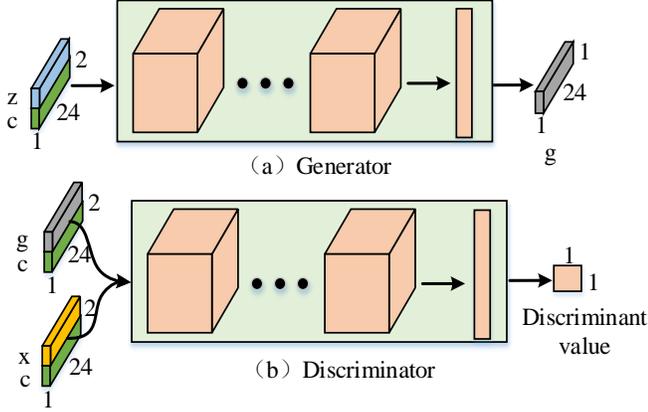

Fig. 7 The network architecture of the G and D

The random noise $z$ is spliced vertically with the conditional information $c$ into a matrix, which is fed into the generator. The real sample $x$ or generated sample $g$ is spliced vertically with the conditional information $c$ into a matrix, which is fed into the discriminator. Then, G and D extract the input information by a multilayer convolutional neural network. Eventually, the full connection layer is used as the output of the G and D to obtain the specified size. The structural details of C-LSGANs are shown in Fig. 8. The Leaky-ReLU activation function is employed in the G and D in addition to the output layer. Meanwhile, the batch normalization is introduced to avoid the parameters entering the saturation or dead zone of the activation function during training, which can enhance the training ability of the network. G and D are trained with the Adam optimizer and the parameters $\beta_1$ and $\beta_2$ are fixed to 0.5 and 0.999, respectively. In summary, the algorithmic flow of C-LSGANs can be found in Algorithm 1.

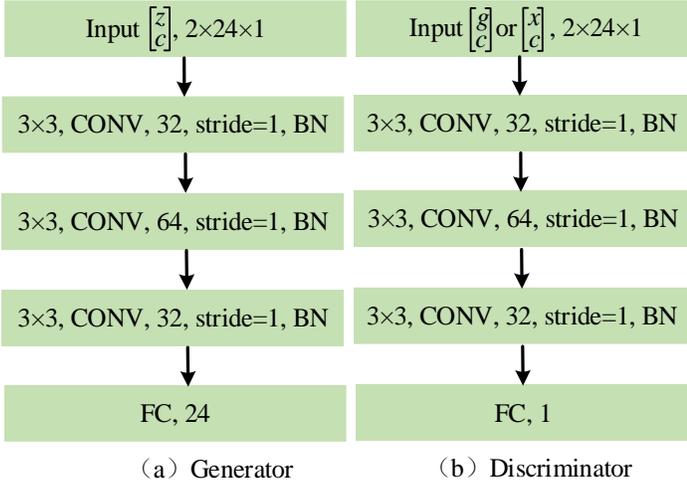

Fig. 8 The C-LSGANs model structure

**Algorithm 1:** C-LSGANs
**Initialize:** $\theta_d, \theta_g, \alpha, \beta_1, \beta_2, m$ – initial C-LSGANs method parameter
1: **for** *episode =1 to E* **do**
    *# Update parameter for Discriminator*
2: Sample batch noise samples
3: Sample batch real samples from training set
4: Sample batch conditional information from training set
5: Update D network:

$$\begin{cases} g_{\theta_d} \leftarrow \nabla_{\theta_d} \frac{1}{m} \sum_{i=1}^{m} \left[ \frac{1}{2}\left(D\left(x^{(i)} \mid c^{(i)}\right)-1\right)^2 + \frac{1}{2}\left(D\left(G\left(z^{(i)} \mid c^{(i)}\right)\right)\right)^2 \right] \\ \theta_d \leftarrow \theta_d - \alpha \cdot Adam\left(\theta_d, g_{\theta_d}, \beta_1, \beta_2\right) \end{cases}$$

    *# Update parameter for Generator*
6: Sample batch noise samples
7: Sample batch conditional information from training set
8: Update G network:

$$\begin{cases} g_{\theta_g} \leftarrow \nabla_{\theta_g} \frac{1}{m} \sum_{i=1}^{m} \left[ \frac{1}{2}\left(D\left(G\left(z^{(i)} \mid c^{(i)}\right)\right)-1\right)^2 \right] \\ \theta_g \leftarrow \theta_g - \alpha \cdot Adam\left(\theta_g, g_{\theta_g}, \beta_1, \beta_2\right) \end{cases}$$

9: **end for**

*4) The evaluation index of the generated scenarios*

The generated scenarios set should accurately capture the inherent characteristics of uncertain factors in RES. Thus, two evaluation indexes are used to measure the quality of the generated scenario set: the coverage rate and the envelope area of the scenario set [51]. The purpose of the former is to assess if the generated scenario data is capable of covering the actual amount of renewable energy generated in each period. The latter aims to evaluate whether the generated scenario data is capable of covering the actual data with the smallest possible area. Noted that, these two evaluation indexes cannot be used alone as evaluation standards and need to be combined. When the coverage rate is larger and the envelope area is smaller, the methodology has much better characteristics. They are shown as follows:

$$\begin{cases} index1 = \frac{1}{N} \sum_{n=1}^{N} B_n \\ B_n = \begin{cases} 1, \text{ if } P_n^{\text{real}} \in [\min\left(P_n^{\text{gene}}\right), \max\left(P_n^{\text{gene}}\right)] \\ 0, \text{ otherwise} \end{cases} \end{cases} \quad (45)$$

$$index2 = \frac{1}{N} \sum_{n=1}^{N} \left(\max\left(P_n^{\text{gene}}\right) - \min\left(P_n^{\text{gene}}\right)\right) \quad (46)$$

## V. NUMERICAL SIMULATION

### A. Case Study Setup

To demonstrate the applicability of the presented approaches, this section carries out a simulation analysis. For the uncertainty of RES, C-LSGANs is used for scenario generation to describe. For the optimal energy allocation problem of MEMG, TD3 is used to solve it. The parameters of the device can be found in Table I. The natural gas price and carbon tax price are set to 0.35\$/m$^3$ and 0.0316\$/kg, respectively. The $CO_2$ intensity of natural gas and electricity purchased from the main network is 0.245 kg/kWh and 0.683 kg/kWh, respectively [8].

TABLE I
THE PARAMETERS OF THE DEVICE

| Parameter | Value | Parameter | Value |
|---|---|---|---|
| $P_{\max}^{\text{WT}}$ | 500kW | $C_{\max}^{\text{ESS}}$ | 1200kW |
| $P_{max}^{PV}$ | 500kW | $\eta_{ch}^{ESS}/\eta_{dis}^{ESS}$ | 0.95/0.95 |
| $P_{e,max}^{CHP}$ | 1200kW | $P_{ch,max}^{ESS}/P_{dis,max}^{ESS}$ | 500/500kW |
| $H_{h,max}^{GB}$ | 2500kW | $C_{max}^{TSS}$ | 800kW |
| $Q_{c,max}^{EC}$ | 4000kW | $\eta_{ch}^{TSS}/\eta_{dis}^{TSS}$ | 0.95/0.95 |
| $Q_{c,max}^{AC}$ | 2500kW | $H_{ch,max}^{TSS}/H_{dis,max}^{TSS}$ | 500/500kW |
| $\lambda^{ESS}$ | 0.999 | $C_{max}^{CSS}$ | 1200kW |
| $\lambda^{TSS}$ | 0.999 | $\eta_{ch}^{CSS}/\eta_{dis}^{CSS}$ | 0.95/0.95 |
| $\lambda^{CSS}$ | 0.999 | $Q_{ch,max}^{CSS}/Q_{dis,max}^{CSS}$ | 500/500kW |

Moreover, the hyperparameters of TD3 can be found in Table II. The algorithms for C-LSGANs and TD3 were coded in Python using Spyder, and the multilayer neural networks used in the algorithms were developed using PyTorch. All tests were performed on a computer with an AMD R7-5800H CPU and 16GB of RAM.

TABLE II
THE HYPERPARAMETERS OF TD3

| Hyperparameter | Value |
|---|---|
| $\alpha_\pi$ | 0.000005 |
| $\alpha_Q$ | 0.00005 |
| $\gamma$ | 0.95 |
| $M$ | 256 |
| $D$ | 36000 |
| $\tau$ | 0.001 |

### B. Comparison of scenario generation methods

1000 curves of wind power and photovoltaic are produced using the trained G, as shown in Fig. 9. In the figure, the generated data can cover the real data well, which shows the validity of the C-LSGANs.

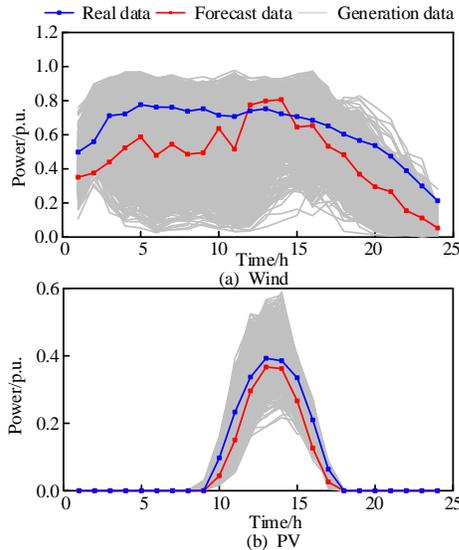

Fig. 9. The generated data of RES

In addition, the C-WGAN-GP and Monte-Carlo methods were chosen to evaluate the behavior of the presented approaches. The 40-day data was employed for testing. 2 evaluation indexes of each method are shown in Table III. For the generated data of wind power, all three methods have a high coverage rate. However, the proposed method has a lower envelope area in this paper, which shows that the description of wind power uncertainty is better. For the generated data of photovoltaic, the C-WGAN-GP method has a lower envelope area, but its coverage rate is also lower. Compared to C-WGAN-GP, the coverage rates of C-LSGANs and the Monte-Carlo method improved by 7.31% and 10.74%, respectively. The envelope area of C-LSGANs is 0.0069 smaller than Monte-Carlo. Hence, the C-LSGANs method is employed to generate data to describe the uncertainty of RES. Finally, the generated data set is used in the later scheduling, and the superior energy supply sector can formulate solutions in advance to cope with the uncertainty of RES.

TABLE III
THE INDEX COMPARISON OF DIFFERENT METHODS

| Methods | | C-LSGANs | C-WGAN-GP | Monte-Carlo |
|---|---|---|---|---|
| Wind | Index1(%) | 92.41 | 90.74 | 92.13 |
| | Index2(%) | 0.4292 | 0.6770 | 0.5519 |
| PV | Index1(%) | 95.74 | 88.43 | 99.17 |
| | Index2(%) | 0.0810 | 0.0788 | 0.0879 |

### C. Discussion of dispatch outcome

#### 1) Algorithm performance evaluation

In this section, the SAC and DDPG algorithms are selected for comparison to verify the validity of the presented approach. The average return convergence curve of the 3 algorithms can be found in Fig. 10. In the figure, DDPG has the worst convergence. Although SAC converges faster than TD3, TD3 can converge to a better reward value. In addition, compared with TD3, SAC needs to update an additional parameter that decides the weight of the entropy versus the reward. Therefore, under the same episode, SAC requires longer training time. The training time comparison is shown in Table IV.

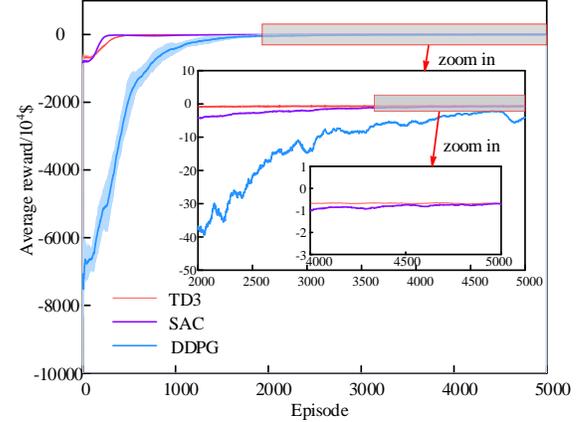

Fig. 10. The average return convergence curve

TABLE IV
THE COMPARISON OF THE TRAINING TIME

| Algorithm | TD3 | SAC |
|---|---|---|
| Training time (s) | 855.56 | 1486.23 |
| Solution time (s) | 0.083 | 0.088 |

#### 2) Analysis of scheduling results
##### a) Cost comparison

Four seasons of spring, summer, autumn, and winter separately select a typical day for research in this paper. The cost comparison of different methods is shown in Table V. It can be seen from the table that the calculation results of TD3 are better than PSO. Compared with PSO, the operating costs of TD3 were decreased by $166.23, $208.32, $163.05, and $351.68, respectively, which demonstrates the validity and superiority of the approach. Compared with MP, the operating costs of TD3 were increased by $780.56, $516.77, $576.25, and $638.56, respectively. However, in terms of solution time, TD3 has obvious advantages over PSO and MP. Using the TD3 method in the study, the optimization strategy can be obtained by executing the actor network after the network training is finished. In contrast, PSO and MP need to be re-optimized when disturbed, which exacerbates the computational burden. Hence, TD3 has certain advantages over PSO and MP in solving the optimization problem. This also proved the validity of the presented approach in solving the optimal energy allocation problem of MEMG.

TABLE V
THE COST COMPARISON OF DIFFERENT METHODS

| Algorithm | Season | Cost ($) |
|---|---|---|
| TD3 | Spring | 6088.71 |
| | Summer | 6568.59 |
| | Autumn | 6651.43 |
| | Winter | 6745.76 |
| PSO | Spring | 6254.94 |
| | Summer | 6776.91 |
| | Autumn | 6814.48 |
| | Winter | 7097.44 |
| MP | Spring | 5308.15 |
| | Summer | 6051.82 |
| | Autumn | 6075.18 |
| | Winter | 6107.20 |

b) Scheduling results

Fig 11 shows the scheduling results for 4 typical days. Each column is electric, thermal, and cold energy, and natural gas balance, respectively. Note that the scheduling results are obtained based on the forecast values of RES. The impact of RES uncertainty is not considered in this section.

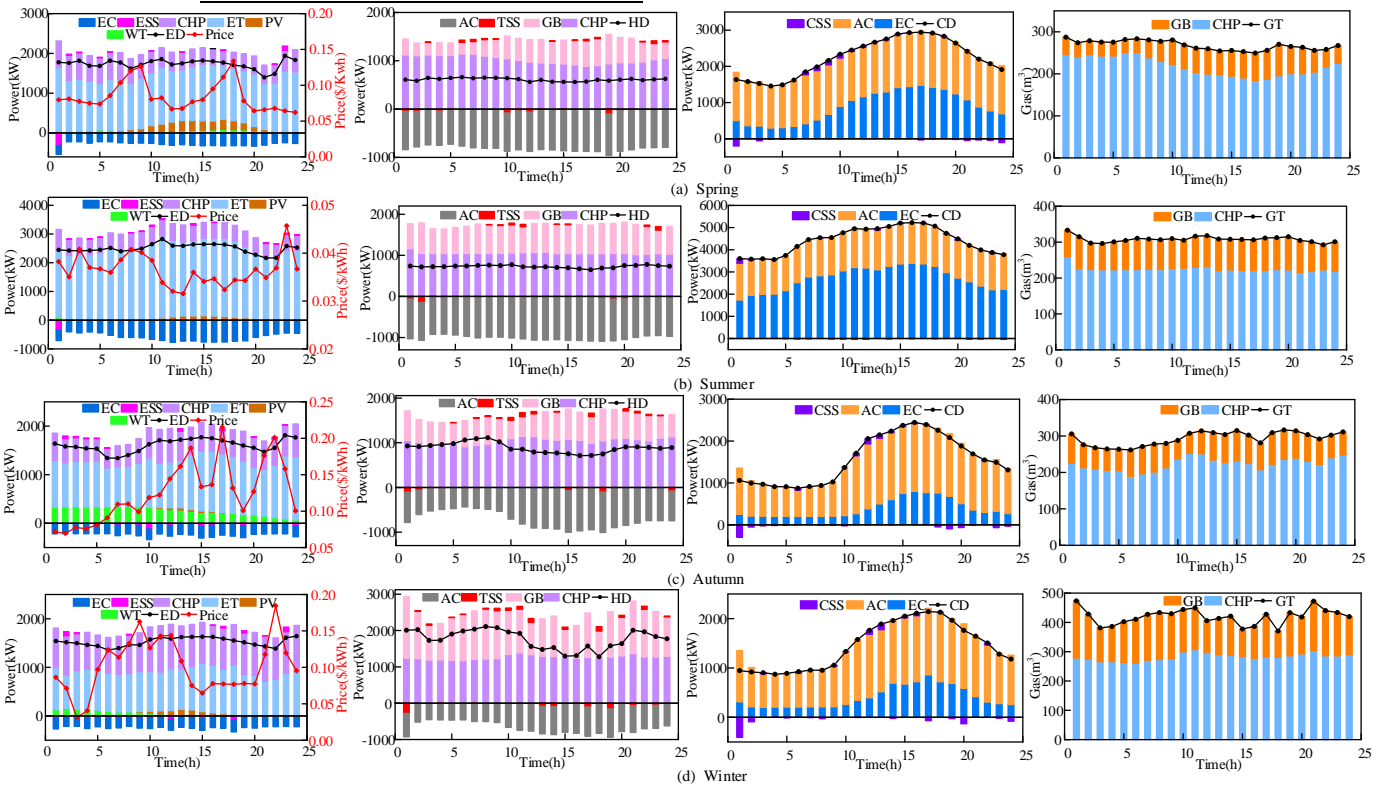

Fig. 11 Dispatch outcomes for 4 typical days

In MEMG, the cooling energy load is met by EC and AC. The heat energy load is supplied by CHP and GB. On a typical day in the summer, there is a lot of demand for cold energy and the price of electricity is relatively cheap. Therefore, EC provides a higher share of cold energy and MEMG has a higher proportion of electric energy purchases. In the typical days of spring and autumn, the electricity price is relatively high. Hence the proportion of cold energy provided by EC is relatively low. This indicates that the agent can make correct decisions based on the electricity price signal.

On a typical day in winter, the thermal energy demand and electricity prices are higher. Since the CHP unit is capable of generating both electrical and thermal energy, it allows the MEMG to meet the thermal energy while reducing the purchasing of electrical energy, which facilitates the economic operation of MEMG. It can be seen from the gas purchase in the last column that CHP consumes a higher proportion of natural gas, which is due to the higher energy conversion efficiency of CHP. This also demonstrates that the presented approach in the study can make appropriate decisions.

*3) Impact Analysis of off-design model*

Fig. 12 is the comparison of operating costs between off-design performance and rated performance.

In Fig. 12, the operating cost of the rated performance is lower. This is because the device is always in optimal operating condition, which allows for maximum energy utilization. Under the off-design performance operation, the operating efficiency of the device changes with the load rate, and it is not possible to ensure that the equipment is at the optimal operating point at all times, so the energy conversion efficiency is lower. This leads to the achievement of the same amount of energy output requiring more energy input, resulting in increased operating costs. Notably, the equipment is not always able to operate at the optimal operating point in practice. Therefore, considering the off-design performance of the device can more truly reflect the operating status of the device, and can also make the scheduling results more accurate.

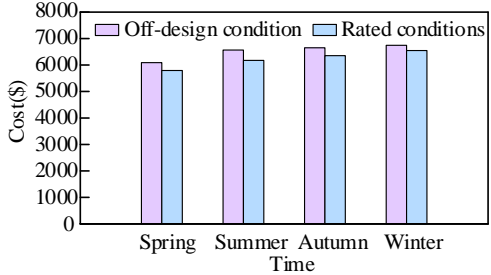

Fig. 12 The cost comparison of different model

*4) Impact Analysis of RES Uncertainty*

To test the ability of DRL agents to cope with short-term uncertainty, the summer typical day emergency simulation was carried out. At 01:00, the wind power output decreases from the maximum output to zero. At 12:00, the wind power output increases sharply. At 16:00, the photovoltaic output decreases from the maximum output to zero. At 19:00, the photovoltaic output increases suddenly. The output of each unit is shown in Fig. 13.

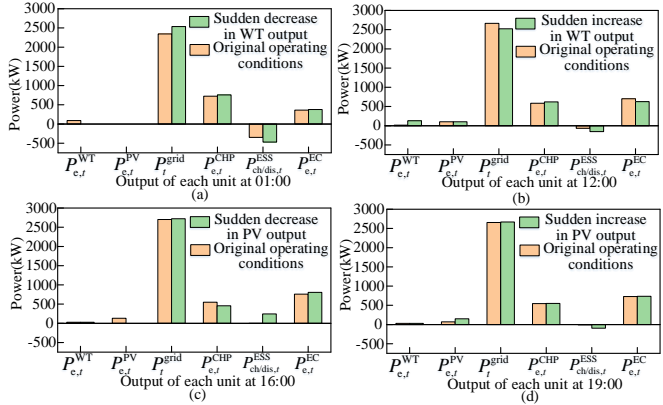

Fig. 13 Scheduling strategy for emergency simulation

When an emergency occurs, the DRL agent can still make reasonable actions to maintain the power balance of the system. At 01:00, WT output decreases, and power purchase and CHP output increase to meet the demand for electric load. At 12:00, WT output increases, power purchase decreases and ESS energy storage power increases. At 16:00, the PV output decreases, and the DRL agent increases the power purchase and ESS energy release power. At 19:00, the PV output increases and the ESS energy storage power increases to maintain the power balance. Therefore, DRL agent can deal with the short-term uncertainty of MEMG.

Figs. 14 and 15 display the intervals of purchased electricity and natural gas of 4 typical days, respectively.

Due to the natural characteristics of RES, there is a certain error in the forecast power, which will affect the electricity and natural gas purchases in MEMG. The uncertainty propagates from the MEMG to the superior energy supply sector, reducing the stability of MEMG. Therefore, to tackle the uncertainty of RES, it is crucial for the superior energy supply sector to formulate countermeasures in advance to guarantee the reliable operation of MEMG. In the study, C-LSGANs-based generated scenarios are used to describe the uncertainty of RES, and the generated data are used to formulate the scheduling strategy to obtain the upper and lower intervals of purchased electricity and natural gas. Based on this, the superior energy supply sector can formulate corresponding solutions to deal with uncertainty.

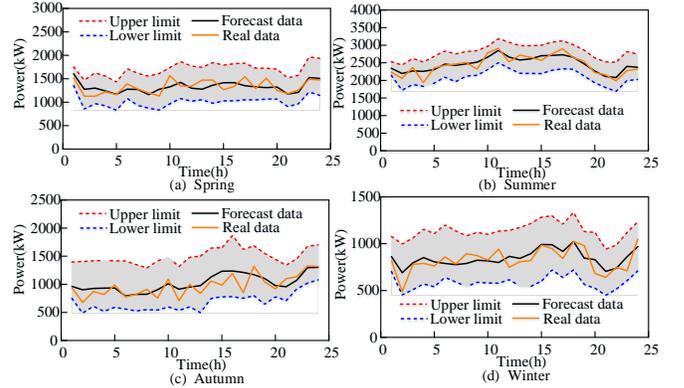

Fig. 14 The intervals of purchased electricity

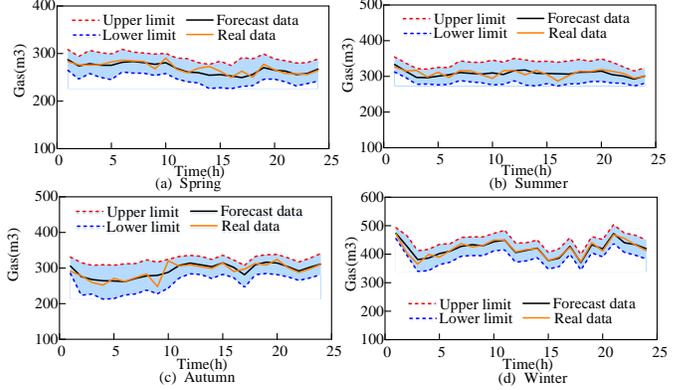

Fig. 15 The intervals of purchased natural gas

## VI. CONCLUSIONS

The MEMG energy management problem considering the off-design characteristics of the device and the uncertainty of RES is investigated in this paper. Case studies are performed to demonstrate the validity of the presented methodology. The main findings are summarized as below:

1) The DRL method is adopted to deal with the nonlinear optimization problem of MEMG with the off-design performance of the device. Compared to rated performance, the off-design performance model is more reflective of the true operating conditions of the device, but the operating costs of MEMG have increased by $296.5, $393.7, $295, and $196.3, respectively.

2) Aiming at the difficulty of describing the uncertainty of RES, the C-LSGANs approach based on RES forecast data is presented to construct the generation scenario of RES. Compared to C-WGAN-GP and Monte-Carlo, the coverage rate of the generated WP scenario is reduced by 1.67% and 0.28%, respectively. Thus, the C-LSGANs method has better results in generating data for RES.

3) The generated data of RES is employed for scheduling to

obtain caps and floors for the purchase of electricity and natural gas. Based on this, the superior energy supply sector can formulate solutions in advance to cope with the uncertainty of RES.

With the increase of the number of MEMGs on the consumer side, different MGs can exchange energy to ensure their economic benefits. Given this, future research will focus on the application of multi-agent DRL in multi-MG systems to achieve privacy protection and improve the economy of the MEMG.

ACKNOWLEDGMENT

This work is supported by National Natural Science Foundation of China (51777027).

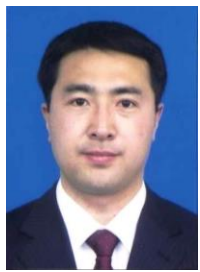
**Yang Cui** (Member, IEEE, Member, CSEE) received the Ph.D. degree from North China Electric Power University, Beijing, China, in 2011. He is a professor in the School of Electrical Engineering, Northeast Electric Power University, Jilin, China. His research interests include coordinated regulation of large-scale renewable energy network consumption, analysis and control of new power system, and the theory and application of integrated energy system optimal scheduling and other directions

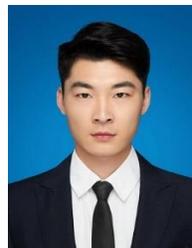
**Yang Xu** (Student Member, IEEE, Member, CSEE) received the B.S. and M.S. degrees from Northeastern Electric Power University, Jilin, China, in 2019 and 2022, respectively. He is currently pursuing his Ph.D. degree at Northeastern Electric Power University, Jilin, China. His research interests include the application of artificial intelligence in energy power systems, electricity market, and optimized operation of integrated energy systems.

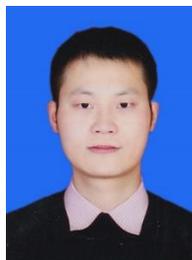
**Yang Li** (Senior Member, IEEE) was born in Nanyang, China. He received his Ph.D. degree in Electrical Engineering from North China Electric Power University (NCEPU), Beijing, China, in 2014.
He is a professor at the School of Electrical Engineering, Northeast Electric Power University, Jilin, China. From Jan. 2017 to Feb. 2019, he was also a postdoc with Argonne National Laboratory, Lemont, United States. His research interests include renewable energy integration, AI-driven power system stability analysis, and integrated energy system. He is featured in Stanford University's List of the World's Top 2% Scientists for the year 2022. He serves as an Associate Editor for the journals of IEEE Transactions on Industrial Informatics, IEEE Transactions on Industry Applications, and IET Renewable Power Generation. He is also a Young Editorial Board Member of Applied Energy.

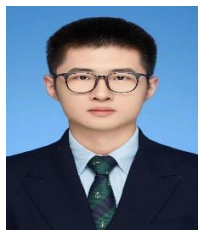
**Yijiian Wang** is currently pursuing a M.S. degree at Northeast Electric Power University His research interests include machine learning, particularly, deep reinforcement learning.

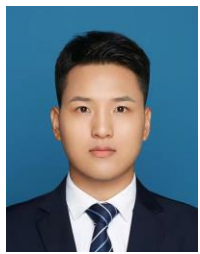
**Xinpeng Zou** received the B.S. degree from Northeastern Electric Power University, Jilin, China, in 2021. He is currently pursuing his M.S. degree at Northeastern Electric Power University, Jilin, China. His research interest is the coordinated dispatching of new energy power generation networking.